\documentstyle[12pt]{article}

\def\GeV{{\rm GeV}}
\def\roughly#1{\raise.3ex
    \hbox{$#1$\kern-.75em\lower1ex\hbox{$\sim$}}}
\def\be{\begin{equation}}
\def\ee{\end{equation}}
\def\bea{\begin{eqnarray}}
\def\eea{\end{eqnarray}}
\def\sw{\sin^2\theta_W}
\def\mz{m_Z}
\def\mg{M_G}
\def\mh{m_{H_2}}
\def\msigma{M_\Sigma}
\def\mtr{M_{\rm tr}}
\def\mcr{M_{\rm tr}^{\rm cr}}
\def\mx{M_{\rm X}}
\def\ainv{\overrightarrow{\alpha}^{-1}}
\def\onevec{\overrightarrow{1}}
\def\epsvec{\overrightarrow{\epsilon}}
\def\deltatwo{\overrightarrow{\Delta}_2}
\def\bvec{\overrightarrow{\beta}}
\def\btr{\overrightarrow{\beta}_{\rm tr}}
\def\pvec{\overrightarrow{P}}
\def\mamg{\left(\frac{M_a}{\mg}\right)}
\def\mp{\hat{M}_P}
\def\tr{\,{\rm tr}\,}
\def\msbar{\rm\overline{MS}}
\def\drbar{\rm\overline{DR}}

\begin{document}
\begin{titlepage}
\begin{center}
October 1992\hfill    LBL-32905 \\
               \hfill    UCB-PTH-92/37 \\
               \hfill    hep-ph/9210240
\vskip .2in

{\large \bf
Gravitational Smearing of Minimal Supersymmetric Unification
Predictions}\footnote{This work
was supported in part by the Director, Office of Energy
Research, Office of High Energy and Nuclear Physics, Division of
High
Energy Physics of the U.S. Department of Energy under Contract
DE-AC03-76SF00098, and in part by NSF grant PHY-90-21139.}

\vskip .4in

Lawrence J. Hall\footnote{
E-mail: hall\_lj@lbl}\\[.15in]

{\em Theoretical Physics Group, 50A/3115\\
      Lawrence Berkeley Laboratory\\
     1 Cyclotron Road\\
     Berkeley, California 94720}

and

{\em Physics Department\\
     University of California\\
     Berkeley, California 94720}

\vskip 16pt
Uri Sarid\footnote{E-mail: sarid@obelix.lbl.gov}\\[.15in]

{\em Theoretical Physics Group, 50A/3115\\
      Lawrence Berkeley Laboratory\\
     1 Cyclotron Road\\
     Berkeley, California 94720}
\end{center}

\vskip .2in

\begin{abstract}

%

The prediction for $\alpha_s$ in the minimal supersymmetric SU(5)
grand
unified
theory is studied in the presence of a gravitationally-induced
dimension-five
operator.  Unless the coefficient of this operator is small, the
correlation between $\alpha_s$ and the mass scale which governs
proton
decay to
K$\nu$ is destroyed. Furthermore, a reduction of the experimental
uncertainty in $\alpha_s$ would not provide a significant test of
the
theory.


\end{abstract}
\end{titlepage}

\renewcommand{\thepage}{\roman{page}}
\setcounter{page}{2}
\mbox{ }

\vskip 1in

\begin{center}
{\bf Disclaimer}
\end{center}

\vskip .2in

\begin{scriptsize}
\begin{quotation}
This document was prepared as an account of work sponsored by the
United
States Government.  To ensure plausible deniability, neither the
United
States Government nor any agency
thereof, nor The Regents of the University of California, nor any of
their
employees, makes any warranty, express or implied, or assumes any
legal
liability or responsibility for the accuracy, completeness, or
usefulness
of any information, apparatus, product, or process disclosed, or
represents
that its use would not infringe privately owned rights.  Reference
herein
to any specific commercial products process, or service by its trade
name,
trademark, manufacturer, or otherwise, does not necessarily
constitute or
imply its endorsement, recommendation, or favoring by the United
States
Government or any agency thereof, or The Regents of the University
of
California.  The views and opinions of authors expressed herein do
not
necessarily state or reflect those of the United States Government
or any
agency thereof of The Regents of the University of California and
shall
not be used for advertising or product endorsement purposes.
\end{quotation}
\end{scriptsize}

\vskip 2in

\begin{center}
\begin{small}
{\em Lawrence Berkeley Laboratory is an equal opportunity employer.}
\end{small}
\end{center}

\newpage
\renewcommand{\thepage}{\arabic{page}}
\setcounter{page}{1}
\setcounter{footnote}{1}

The standard model of elementary particles and interactions is
specified
by eighteen fundamental parameters. The only one which has been
successfully predicted to a high level of accuracy
is the weak mixing angle. In simple supersymmetric versions of grand
unified
theories (GUTs) it is predicted \cite{firstones} to be $0.233$, in
full
agreement with its experimental value \cite{expinputs} of
$0.2325\pm0.0008$. Over the past few years several groups
\cite{general,ghs,yanagida,hagiwara} have studied what information
may be
extracted from this unification of gauge couplings. It is frequently
stated that to further test GUTs, the strong coupling should be
measured
more precisely \cite{glashow,yanagida}. Some claim that this would
determine the scale of superpartner masses, others that
such improved accuracy would help pin down the proton decay rate in
the
minimal model \cite{yanagida}.

The renormalisation group equations for the gauge couplings provide
three
equations relating several quantities. One of
these equations determines the GUT gauge coupling to be about 1/25,
but
beyond this fact it will
not concern us further. A second provides a relationship among the
various
superheavy particle masses. The third predicts one of the low energy
gauge
couplings; although this is frequently chosen to be the weak mixing
angle, we choose instead \cite{ghs} to predict the strong coupling
$\alpha_s$, because we
want to address the question of whether more information would be
gained
by reducing
the error bar on its experimental value.

In a general GUT the spectrum of superheavy states is likely to be
complicated.
Logarithmic threshold corrections to the prediction for $\alpha_s$
may
arise from each nondegenerate (``split'') SU(5) multiplet, and some
such corrections will be present in every GUT \cite{bh}.
The minimal number of SU(5) representations whose states are not
degenerate is
two: one contains the superheavy (mass $\mx$) and the light gauge
particles, the other contains the superheavy (mass $\mtr$) and the
light
members of the multiplet responsible for spontaneous electroweak
symmetry
breaking. In addition, the
remnants $\Sigma$ of the representation which breaks SU(5) to $\rm
SU(3)\times SU(2)\times U(1)$ have masses $\msigma$ which
generically
differ from those of the Goldstone bosons eaten by the superheavy
gauge
particles. It is well known that (at one-loop order) the prediction
for
$\alpha_s$ does not depend on $\mx$. Recently it was pointed out
\cite{hagiwara,yanagida} that there
is also no dependence on $\msigma$ --- a fact overlooked in the
previous
analysis of threshold corrections \cite{bh}. This raises the
interesting
possibility that,
in certain simple GUT models, the only significant dependence of
the $\alpha_s$ prediction on the superheavy sector is through the
mass
parameter
$\mtr$, which in these models controls the rate of proton decay.
Since
$\alpha_s$ increases with
$\mtr$, an improved experimental upper limit on $\alpha_s$ could
reduce the upper bound on $\mtr$. In that case,
super-Kamiokande could definitively test this theory
\cite{yanagida}.

In this letter we study the extent to which these claims are
spoiled by the presence of higher-dimensional operators
generated at the Planck scale \cite{hill,shawet,drees}. Although we
study
the minimal supersymmetric
SU(5) theory, for a wide class of GUTs the superheavy corrections
are at
least as large as those we consider.
We assume that the mass scale suppressing such nonrenormalizable
operators
is the {\em reduced} Planck mass
$\mp\equiv(8\pi\mbox{G}_{\mbox{\scriptsize
N}})^{-1/2}\simeq2.4\times10^{18}\,\GeV$ since this is the
combination
which enters quantum-gravity calculations.
At first sight the Planck scale corrections appear to change the
prediction for $\alpha_s$ only at the 1\% level, which is the ratio
of the
naive GUT scale (roughly $\mx$) to the reduced Planck mass.
But $\mx$ may easily be higher than the naive expectation if
$\msigma$ is
reduced. Furthermore, these corrections are enhanced by several
numerical
factors. As we show below, the result is a significant modification
to the
prediction of $\alpha_s$, and since the sign of this effect is
unknown,
the prediction is spread out considerably.

To one-loop order, and {\em in the absence of gravitational
corrections},
gauge coupling unification is embodied in the three
renormalization-group
equations relating the values of the gauge couplings at the Z mass,
$\ainv
\equiv \ainv(\mz) = (\alpha_1^{-1},\alpha_2^{-1},\alpha_3^{-1})$,
and the
common gauge coupling $\alpha_G$ at the GUT scale $\mg$:
\be
\ainv = \alpha_G^{-1} \onevec - \sum_a \bvec_a \ln\mamg
\label{gauni}
\ee
Here $\onevec\equiv (1,1,1)$ and $\bvec_a\equiv \vec{b}_a/(2\pi)$
where
$\vec{b}_a$ are the three beta-function coefficients for the
particle
labeled by $a$.
The sum extends over all particles in the model, and $M_a$ denotes
the
mass threshold at which each is integrated out. (We neglect
electroweak-breaking effects in the SUSY mass spectrum, and treat
the top
quark as being degenerate with the Z. These effects are small
relative to
the dominant uncertainties in the experimental inputs and in the
gravitational corrections.)  All of the standard model particles
are already present at the initial scale $\mz$; we then include the
second
Higgs doublet at $\mh$,  the squarks at an average mass
$m_{\tilde{q}}$,
the sleptons at their average mass $m_{\tilde{l}}$, the winos at
$m_{\tilde{w}}$, the gluinos at $m_{\tilde{g}}$, the higgsinos at
$m_{\tilde{H}}$, the color-triplet component of the {\bf 5} of Higgs
at
$\mtr$, the non-Goldstone-boson members of the {\bf 24} of Higgs at
$\msigma$, and finally the superheavy gauge bosons and their
superpartners
(`X') at $\mx$.
The GUT scale is the highest mass threshold, above which all
particles
fill complete SU(5) multiplets.
The experimental inputs are derived from \cite{expinputs}
$s^2\equiv\sw =
0.2325 \pm 0.0008$ and $1/\alpha = 127.9 \pm 0.1$ using
$\ainv = (\frac{3}{5}(1-s^2)/\alpha,
s^2/\alpha,1/\alpha_{\rm s})$.
The two-loop contributions to (\ref{gauni}) are incorporated using
typical
values for all the parameters. We also include in these
contributions the
terms needed to translate the experimental inputs given in the
$\msbar$
scheme into the $\drbar$ scheme appropriate for step-function
corrections
in supersymmetric theories \cite{drref}. The result is a term
$\deltatwo =
(0.65,1.09+2/12\pi,0.55+3/12\pi)$ which must be added to the
right-hand
side of (\ref{gauni}).

We concentrate on the predictions of (\ref{gauni}) for $\alpha_s$
and
$\mtr$, and ignore the prediction for $\alpha_G$. Therefore we
consider
(well-known) linear combinations of (\ref{gauni}) which do not
involve
$\alpha_G$.
Define for convenience (up to an irrelevant overall normalization)
one
projection vector $\pvec_1$ by requiring $\pvec_1\cdot\onevec=0$ and
$\pvec_1\cdot\bvec_X = 0$, and another projection vector $\pvec_2$
by
requiring $\pvec_2\cdot\onevec=0$ and $\pvec_2\cdot\btr = 0$. We
choose
\be
\pvec_1 = (-1,3,-2);\qquad\pvec_2 = (5,-3,-2). \label{pi0}
\ee
The dot product of $\pvec_1$ with (\ref{gauni}) will be independent
of
$\alpha_G$ and of $\mx$ and $\msigma$, and therefore will furnish a
simple
expression for $\alpha_s$ in terms of the low-energy parameters and
the
Higgs-triplet mass. The dot product of $\pvec_2$ with (\ref{gauni})
will
be independent of $\alpha_G$ and of $\mtr$ and the light Higgs
sector, and
will relate the masses of the superheavy gauge multiplet and the
superheavy {\bf 24}. Finally, the unification scale $\mg$ enters
(\ref{gauni}) only through the combination
$(\ln\mg)\,{\displaystyle\sum_a} \bvec_a \propto \onevec$ and so it,
too,
is projected out; thus any other scale may be used in these dot
products,
and we  choose that scale for convenience to be $\mz$.

Note that in the dot product with $\pvec_1$ both the $X$ {\em and}
the
$\Sigma$ were projected out. The reason is simple.
The $\bvec$ for any complete SU(5) multiplet, and in particular
$\bvec_{\bf 24}$ for the {\bf 24}, contributes equally to the
running of
all gauge couplings, so it is proportional to $\onevec$ and hence
orthogonal to $\pvec_1$. The $\bvec_{\rm GB}$ for the Goldstone mode
components of the {\bf 24} is proportional to the $\bvec_X$ of the
superheavy X since they carry the same quantum numbers, and so it
too is
orthogonal to $\pvec_1$. Therefore their difference $\bvec_{\bf 24}
-
\bvec_{GB} = \bvec_{\Sigma}$
also satisfies $\pvec_1\cdot\bvec_\Sigma=0$ and the $\Sigma$ does
{\em
not} make a threshold contribution to this equation.
Similarly, both the Higgs doublet {\em and} the triplet are
projected out
in the dot product with $\pvec_2$. The $\Sigma$ could contribute if
it
were split, which is not the case in the minimal model but would be
the
case in most extensions. An example of such a contribution in the
minimal
model is provided by the gauginos: they would also be projected out
since
they carry quantum numbers complementary to those of the $X$, but
their
masses are widely split by renormalization-group running so they
make a
significant (and calculable) contribution to the predictions for
$\alpha_s$.

To obtain specific predictions, we need the mass spectrum of the
model. In
the minimal model the weak-scale masses are determined to a good
approximation
by the four mass parameters $m_{0}$ (the common scalar mass),
$m_{1/2}$
(the common gaugino mass at the GUT scale), $\mu$ (the coupling of
the two
Higgs doublets in the superpotential) and $\mh$ (the mass of the
second
Higgs doublet---we take the first to be degenerate with the Z). For
our
purposes the following simplified spectrum will suffice
\cite{susyspectrum,yanagida}:
$m_{\tilde{q}}\simeq\sqrt{m_{0}^2+6 m_{1/2}^2}$,
$m_{\tilde{l}}\simeq\sqrt{m_{0}^2+.4 m_{1/2}^2}$,
$m_{\tilde{g}}\simeq 2.7 m_{1/2}$,
$m_{\tilde{w}}\simeq 0.8 m_{1/2}$ and $m_{\tilde{H}}\simeq\mu$. By
applying the projections (\ref{pi0}) to (\ref{gauni}) and including
the
two-loop term $\deltatwo$ we find
\be
{2\over \alpha_s} + {6\over 5\pi}\ln{\mtr\over\mz} =
f_1(s^2,m_{0},m_{1/2},\mu,\mh) \label{spred0}
\ee
and
\be
{2\over \alpha_s} + {6\over\pi}\ln{\msigma\over\mz} +
{12\over\pi}\ln{\mx\over\mz}=
f_2(s^2,m_{0},m_{1/2}) \label{xpred0}
\ee
where
\bea
f_1 &=& {3(6s^2-1)\over5\alpha} -
{3\over20\pi}\ln{m_{0}^2+6m_{1/2}^2\over m_{0}^2+.4m_{1/2}^2} -
{2\over\pi}\ln{2.7\over0.8} +
{4\over5\pi}\ln{\mu\over\mz} +
{1\over5\pi}\ln{\mh\over\mz} - 1.52 \nonumber \\
&\simeq& 27.9 +
0.4\sigma+{1\over\pi}\ln{\mu_{\vphantom{H_2}}^{4/5}
\mh^{1/5}\over\mz},
\label{eqf1} \\
f_2 &=& {3(1-2s^2)\over\alpha} -
{3\over4\pi}\ln{m_{0}^2+6m_{1/2}^2\over m_{0}^2+.4m_{1/2}^2} -
{2\over\pi}\ln(2.7\cdot0.8) -
{4\over\pi}\ln{m_{1/2}\over\mz} + 1.13 + {1\over\pi}
\nonumber \\
&\simeq& 206.2 - 0.6\sigma
- {3\over4\pi}\ln{m_{0}^2+6m_{1/2}^2\over m_{0}^2+.4m_{1/2}^2}
- {4\over\pi}\ln{m_{1/2}\over\mz},
\label{eqf2}
\eea
and
$\sigma\equiv (s^2-0.2325)/0.0008$.

Eq.~(\ref{xpred0}) can be viewed as a prediction for $\msigma$. It
shows
that one can raise the X mass by lowering the mass of the $\Sigma$
without
affecting the prediction for $\alpha_s$. Since there are no
experimental
consequences of a light $\Sigma$, we focus on (\ref{spred0}). To
avoid
excessive fine-tuning and retain the motivation for supersymmetric
unification, we restrict $m_0$, $m_{1/2}$, $\mu$ and $\mh$ to lie
below
one TeV. (Our results are not sensitive to the exact value of this
cutoff.) By varying these four parameters between the weak scale and
a
TeV, and varying $s^2$ within one standard deviation of its central
value
(namely $|\sigma|\le 1$), we obtain a predicted range of $\alpha_s$
as a
function of $\mtr$. This range is the region between the two black
curves
in Fig.~1.

We now turn to the effects of quantum gravity on these predictions.
In the
absence of a specific and predictive theory of quantum gravity, we
can
only estimate these effects by including higher-dimension operators
that
would arise in an effective theory once the Planck-scale degrees of
freedom have been integrated out. Following Hill \cite{hill} and
Shafi and
Wetterich \cite{shawet}, we restrict our attention to the dominant
dimension-five operator
\be
\delta{\cal L} = \frac{c}{2 \mp} \tr(GG\Sigma)\,,\label{odef}
\ee
where $G\equiv G_aT^a$ is the field-strength tensor of the SU(5)
gauge
field and the generators are normalized to $\tr\,
T^aT^b=\frac{1}{2}\delta^{ab}$. The mass scale suppressing this
operator
is the reduced Planck mass $\mp\simeq2.4\times10^{18}\,\GeV$ as
discussed
above. We have also conservatively included a factor of $1/2$ to
account
for the two identical operators $GG$. The remaining coefficient $c$
is
unknown without further assumptions; we have no reason to think it
is less
than ${\cal O}(1)$ in magnitude.
SU(5) is broken by the vacuum expectation value
$\langle\Sigma\rangle\equiv vT^0$ where $T^0 =
\mbox{diag}(2,2,2,-3,-3)\,/2\sqrt{15}\,$. This breaking induces
superheavy
masses $\mx = \sqrt{5\over6}g_5v$ from the covariant derivative
$D_\mu\Sigma = \partial_\mu\Sigma_aT^a+ig_5X_{\mu
a}\Sigma_b[T^a,T^b]$,
a triplet mass $\mtr = \sqrt{5\over12}\lambda_5v$ from the term
${3\over5}\mtr\overline{H}_5H_5 + \lambda_5\overline{H}_5\Sigma H_5$
(after a fine-tuning to make the doublets light),
and a $\Sigma$ mass $\msigma =
{1\over2}\sqrt{5\over12}\lambda_{24}v$ from
the term
${1\over5}\mtr\tr\Sigma^2+\frac{1}{3}\lambda_{24}\tr\Sigma^3$. It
also
modifies the kinetic terms of the standard-model gauge bosons
through
$\delta{\cal L}$:
\bea
{\cal L}_{\mbox{\scriptsize gauge}} =
-{1\over4}(FF)_{\mbox{\scriptsize U(1)}}
\left[1+{c\over\vphantom{\mp}2}{v\over\mp}
\left({-1\over2\sqrt{15}}\right)\right]
& \! - \! &
{1\over2}\tr(GG)_{\mbox{\scriptsize SU(2)}
}\left[1+{c\over\vphantom{\mp}2}{v\over\mp}
\left({-3\over2\sqrt{15}}\right)\right]
\nonumber\\
&  &\hspace{-1.5in}
-\,\,{1\over2}\tr(GG)_{\mbox{\scriptsize SU(3)}} \left[1 +
{c\over\vphantom{\mp}2}
{v\over\mp}\left({1\over\sqrt{15}}\right)\right]\,.
\label{gaumod}
\eea
Consequently the three gauge couplings are not degenerate at the GUT
scale. The first term in the equations for unification (\ref{gauni})
must
be replaced by $\alpha_G^{-1} \onevec \longrightarrow \alpha_G^{-1}
(\onevec+\epsvec)$ where $\epsvec \equiv  (cv/2\mp)
\left(
\frac{-1}{2\sqrt{15}},
\frac{-3}{2\sqrt{15}},\frac{1}{\sqrt{15}}
\right)$. We have absorbed the sign of $v$ into $c$, so $v$ and
$\lambda_{5,24}$ are by definition positive.

By applying the
projection operators to the modified unification equations, we
obtain:
\be
{2\over \alpha_s} + {6\over 5\pi}\ln{\mtr\over\mz}
- \sqrt{12\over5}{c\over2}{v\over\mp}{1\over\alpha_G} =
f_1(s^2,m_{0},m_{1/2},\mu,\mh) \label{spred1}
\ee
and
\be
{2\over \alpha_s} + {9\over\pi}\ln{5\over12} +
{12\over\pi}\ln g_5 +
{6\over\pi}\ln\lambda_{24} +
{18\over\pi}\ln{v\over\mz}=
f_2(s^2,m_{0},m_{1/2})\,. \label{xpred1}
\ee
In (\ref{spred1}) we use
the zeroth-order expression for $\alpha_G=g_5^2/4\pi\simeq1/25$ in
the
coefficient of $\epsvec$. Eq.~(\ref{xpred1}) has no direct
gravitational
contributions, since it turns out that $\pvec_2\cdot\epsvec=0$;
we have merely rewritten it using the above expressions for the
superheavy
masses.
The magnitude of the gravitational smearing may be readily estimated
from
(\ref{spred1}). If $v\sim2\times10^{17}\,\GeV$ and $|c|\sim1$ then
the
prediction for $\alpha_s$ is corrected by $\sim10\%$.

\renewcommand{\thefootnote}{\fnsymbol{footnote}}
To be more precise, we study the predictions of the two equations
(\ref{spred1}) and (\ref{xpred1}) in the five unknowns
$\{\alpha_s,\mtr,\lambda_5=\sqrt{12\over5}\mtr/v,\lambda_{24},c\}$.
A nonzero $c$ couples the two equations and makes an exact analytic
solution
impossible. Instead, they may be solved analytically to a good
approximation ($\sim\pm1.5\%$ in $\alpha_s$ and $\sim\pm30\%$ in
$\mtr$),
or numerically to a high precision. The analytic expressions
are\footnote{The coefficient of the last term in (\ref{sapprox})
should be
changed from $-0.025$ to $-0.04$ for large values (0.14-0.15) of
$\alpha_s$ in order to achieve the desired accuracy.}
\bea
\alpha_s & \simeq & 0.132
\left(1-0.024\sigma-0.02\ln{\mu^{4/5}\mh^{1/5}\over\mz}
+0.025\ln{\mtr\over{3\times10^{16}\,\GeV}} \right. \nonumber \\
 & \phantom{\simeq} & \phantom{ 0.132xxx}\left.
\vphantom{{\mu^{4/5}\mh^{1/5}\over\mz}} -
0.025\,c\,(1-0.1\sigma)\left(m_{1/2}\over\mz\right)^{-2/9}
\lambda_{24}^{-1/3}
\right) \label{sapprox}
\eea
and
\be
\mtr\simeq
(3\times10^{16}\,\GeV)\,
\lambda_5\,(1-0.1\sigma)\,\lambda_{24}^{-1/3}
\left(m_{1/2}\over\mz\right)^{-2/9}.\label{mapprox}
\ee
Numerically, one subtracts (\ref{xpred1}) from (\ref{spred1}) and
solves
the resulting equation for $\mtr$:
\be
f_1-f_2+{6\over\pi}\ln{\lambda_{24}\over\lambda_5^3}+
{6\over\pi}\ln4\pi\alpha_G =
-{84\over5\pi}\ln{\mtr\over\mz}-
{6c\over5\alpha_G\lambda_5}{\mtr\over\mp}.
\label{esimul}
\ee
The solution(s) can then be used to find the corresponding
$\alpha_s$.
Since we have no reason to presume that the scalar couplings
$\lambda_{5,24}$ are particularly small or particularly large, we
allow
them to vary between 0.1 and 3, and also let $c$ vary between $-1$
and
$1$.  For each such choice of $\lambda_5$, $\lambda_{24}$ and $c$,
we
obtain numerically a region of allowed $(\alpha_s,\mtr)$ values when
we
scan $m_0$, $m_{1/2}$, $\mu$, $\mh$ and $s^2$ over the same ranges
as
before. The overlap of all these regions is shown as the gray area
in
Fig.~1; it represents the region allowed in the minimal
supersymmetric
\cite{nonsusy} SU(5) model by our present knowledge of $s^2$, our
suspicions about the ranges of superpartner masses, our assumptions
about
the scalar couplings in the superpotential \cite{maxlambda}, and our
ignorance of the true theory at the GUT scale. The domain of
predictions
for $\alpha_s$ is greatly increased by the possible Planck-scale
corrections, and the correlation between $\alpha_s$ and the
parameter
$\mtr$ relevant to proton decay is largely blurred away.

\section*{Figure Captions}

\begin{description}

\item[Fig.~1:] The prediction in the minimal supersymmetric SU(5)
model of
$\alpha_s$ as a function of the color-triplet mass $\mtr$. The
region
between the two black curves accounts for the possible variation of
the
light superpartner masses and the second Higgs doublet mass between
100
GeV and 1 TeV, and for the variation of $\sw$ between $0.2317$ and
$0.2333$, but does not incorporate any Planck-scale corrections. The
shaded region adds the gravitational corrections, with the
restrictions
that
$0.1\le\lambda_{5,24}\le3$ and that $|c|\le1$.

\end{description}


\begin{thebibliography}{99}

\bibitem{firstones} H. Georgi and S.L. Glashow, Phys. Rev. Lett. 32,
438
(1974); H. Georgi, H.R. Quinn and S. Weinberg, Phys. Rev. Lett. 33,
451
(1974); S. Dimopoulos, S. Raby and F. Wilczek, Phys. Rev. D24, 1681
(1981); S. Dimopoulos and H. Georgi, Nucl. Phys. B193, 150 (1981);
L.
Iba\~{n}ez and G.G. Ross, Phys. Lett. B105, 439 (1981).

\bibitem{expinputs} G. Degrassi, S. Fanchiotti and A. Sirlin, Nucl.
Phys.
B351, 49 (1991); K. Hikasa {\it et al.} (Particle Data Group), Phys.
Rev.
D45
(1992) S1.

\bibitem{general} U. Amaldi, W. de Boer and H. F\"urstenau, Phys.
Lett.
B260, 447 (1991); R. Arnowitt and P. Nath, Phys. Rev. Lett. 69, 725
(1992); J. Ellis, S. Kelley and D.V. Nanopoulos, Phys. Lett. B287,
95
(1992); A.E. Faraggi, B. Grinstein and S. Meshkov, SSC preprint
SSCL-PREPRINT-126-REV (1992); P. Langacker and M. Luo, Phys. Rev.
D44, 817
(1991); G.G. Ross and R.G. Roberts, Nucl. Phys. B377, 571 (1992); A.
Zichichi, CERN preprint CERN-PPE/92-149; and references therein.

\bibitem{ghs} A. Giveon, L.J. Hall and U. Sarid, Phys. Lett. B271,
138
(1991).

\bibitem{yanagida} J. Hisano, H. Murayama and T. Yanagida, Phys.
Rev.
Lett. 69, 1014 (1992); Tohoku University preprint TU-400 (1992),
bulletin
board submission hep-ph 9207279.

\bibitem{hagiwara} K. Hagiwara and Y. Yamada, preprint KEK-TH-331
(1992).

\bibitem{glashow} S.L. Glashow, talk presented at the 1991 joint
Lepton-Photon and EPS Conference, Geneva.

\bibitem{bh} R. Barbieri and L.J. Hall, Nucl. Phys. B364, 27 (1991).

\bibitem{hill} C.T. Hill, Phys. Lett. 135 (1984) 47.

\bibitem{shawet} Q. Shafi and C. Wetterlich, Phys. Rev. Lett. 52
(875)
1984.

\bibitem{drees} M. Drees,  Phys. Lett. B158, 409 (1985) and Phys.
Rev.
D33, 1468 (1986) has also considered nonrenormalizable operators in
SU(5),
showing that the unification scale can be raised to the (reduced)
Planck
mass; however, as he notes, in that case the prediction of the weak
mixing
angle is lost and with it much of the motivation for considering
supersymmetric SU(5) to begin with.

\bibitem{drref} I. Antoniadis, C. Kounnas and K. Tamvakis, Phys.
Lett.
119B, 377 (1982) and references therein.

\bibitem{susyspectrum} See, for example, G.F. Giudice and G.
Ridolfi, Z.
Phys. C 41, 447 (1988).

\bibitem{nonsusy} We note in passing that the non-supersymmetric
minimal
SU(5) model cannot be resurrected by these gravitational effects.
Eqs.~(\ref{spred1}) and (\ref{xpred1}) can easily be rewritten for
this
simpler model using the appropriate beta-function coefficients and
mass
spectrum, yielding expressions for $\alpha_s$ and for $\msigma$. The
first
such expression shows that to obtain $\alpha_S>0.10$ one needs large
gravitational corrections, which require $v>10^{16}\GeV$ for any
plausible
choice of $\sw$. The second expression then shows that $\msigma$ is
proportional to $v^{-22}$ (compare with $v^{-3}$ in the
supersymmetric
version); if $v$ is increased from its unperturbed value of $\sim
10^{14}\GeV$ to the extent required by $\alpha_S$ then $\msigma$ is
reduced to intolerably low values. Of course, if $v$ is at or above
$\mp$
then both (\ref{spred1}) and (\ref{xpred1}) are completely changed,
and
$\msigma$ can be made reasonably large, but in that case we are no
longer
considering an effective SU(5) model.

\bibitem{maxlambda} How large can the gravitational smearing become
if we
relax the lower bound of $0.1$ on $\lambda_{24}$? For $c\ge0$ there
is
always a (unique) solution to (\ref{esimul}), so one could lower the
prediction for $\alpha_s$ as far as desired by lowering
$\lambda_{24}$.
However, for $c<0$ there may be 2, 1 or 0 solutions. For large
$\lambda_{24}$ two solutions exist, one above $\mcr \equiv
-14\lambda_5\alpha_G\mp/\pi c$ and one below. (We have chosen the
lower
one in the above estimates and in the figure.) These two merge into
$\mcr$
at some critical value of $\lambda_{24}$ which depends on the other
parameters, while for smaller $\lambda_{24}$ no solutions exist.
Thus the
largest corrections occur when $\mtr = \mcr$, in which case the
gravitational correction term in (\ref{spred1}) becomes exactly
$84/5\pi$.
Such a correction raises even the lowest possible uncorrected
$\alpha_s$
value, namely $\sim0.118$, to above $0.17$. We learn that any
conceivable
future experimental determination of $\alpha_s$ may be accommodated
using
some value of $\lambda_{24}$. We find that $\lambda_{24}$ need never
be
taken less than $0.015$ for such large corrections, while the
corresponding value of $\mtr$ can be adjusted by changing
$\lambda_5$,
thus allowing the entire experimentally-relevant region in the
$(\alpha_s,\mtr)$ plane.

\bibitem{proofnote} After this work was completed, we learned of a
paper
by P. Langacker and N. Polonsky (Pennsylvania preprint UPR-0513T) in
which
the effects of nonrenormalizable operators arising at the Planck
scale
have also been studied.

\end{thebibliography}
\end{document}